\documentclass[sigconf]{acmart} 
\AtBeginDocument{%
  }

\copyrightyear{2026}
\acmYear{2026}
\setcopyright{cc}
\setcctype{by-nc-nd}
\acmConference[CHI EA '26]{Extended Abstracts of the 2026 CHI Conference on Human Factors in Computing Systems}{April 13--17, 2026}{Barcelona, Spain}
\acmBooktitle{Extended Abstracts of the 2026 CHI Conference on Human Factors in Computing Systems (CHI EA '26), April 13--17, 2026, Barcelona, Spain}
\acmDOI{10.1145/3772363.3798427}
\acmISBN{979-8-4007-2281-3/2026/04}

\usepackage{subcaption}
\usepackage{diagbox}
\usepackage{hhline}
\usepackage{multirow}
\usepackage{colortbl}
\usepackage{graphicx}

\usepackage{tikz}
\usepackage{enumitem}
\usetikzlibrary{shapes.geometric, arrows, positioning, fit}

\begin{document}

\title{GazeCode: Recall-Based Verification for Higher-Quality In-the-Wild Mobile Gaze Data Collection}

\author{Yaxiong Lei}
\email{yl212@st-andrews.ac.uk}
\orcid{0000-0002-0697-7942}
\affiliation{%
  \institution{University of St Andrews}
  \city{St Andrews}
  \country{UK}
}
\affiliation{%
  \institution{University of Essex}
  \city{Colchester}
  \country{UK}
}

\author{Thomas Davies}
\email{td51@st-andrews.ac.uk}
\orcid{0009-0006-9913-5010}
\affiliation{%
  \institution{University of St Andrews}
  \city{St Andrews}
  \country{UK}
}

\author{Xinya Gong}
\email{xg31@st-andrews.ac.uk}
\orcid{0009-0005-6414-9351}
\affiliation{%
  \institution{University of St Andrews}
  \city{St Andrews}
  \country{UK}
}

\author{Shijing He}
\email{shijing.he@kcl.ac.uk}
\orcid{0000-0003-3697-0706}
\affiliation{%
 \institution{King's College London}
 \city{London}
 \country{UK}
}

\author{Juan Ye}
\email{Juan.Ye@st-andrews.ac.uk}
\orcid{0000-0002-2838-6836}
\affiliation{%
 \institution{University of St Andrews}
  \city{St Andrews}
  \country{UK}
}

\renewcommand{\shortauthors}{Yaxiong Lei et al.}

\begin{abstract}
Large-scale mobile gaze estimation relies on in-the-wild datasets, yet unsupervised collection makes it difficult to verify whether participants truly foveate logged targets. Prior mobile protocols often use low-entropy validation (e.g., binary probes) that can be satisfied by guessing and may still allow peripheral viewing, introducing label noise. We present \textbf{GazeCode}, a recall-based verification paradigm for higher-confidence in-the-wild mobile gaze data collection that strengthens \emph{label validity} through a multi-digit recall task (reducing random success to $10^{-N}$) paired with anti-peripheral stimulus design (small, low-contrast, brief digits). The system logs synchronized front-camera video, IMU streams, and target events using high-resolution timestamps. In a formative study (N=3), we probe key parameters (opacity, duration) and directly test peripheral exploitability using an eccentricity-controlled \textit{RING} condition. Results show that low-opacity digits substantially reduce peripheral readability while remaining usable for attentive foveation, supporting the inference that correct recall corresponds to higher-confidence gaze labels. We conclude with actionable design guidelines for robust in-the-wild gaze data collection.
\end{abstract}





\begin{teaserfigure}
  \includegraphics[width=\textwidth]{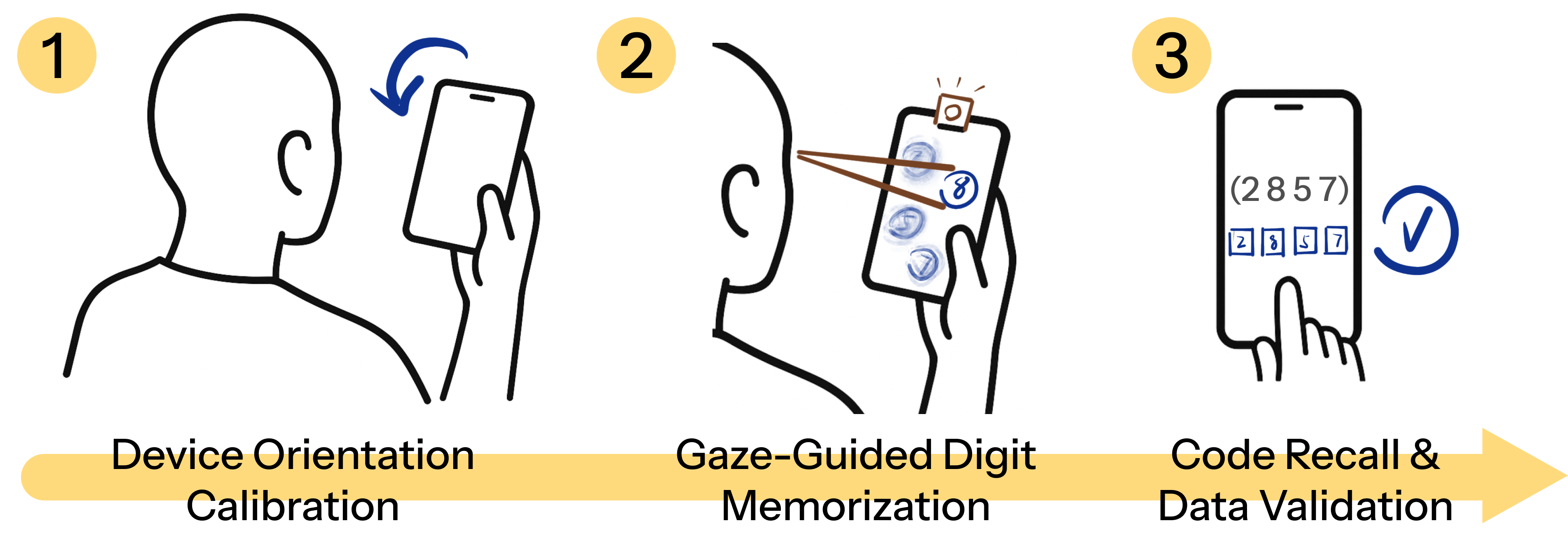}
  \caption{Overview of the GazeCode workflow. (1) Users first calibrate device orientation to ensure diverse gaze angles. (2) During the task, low-contrast digits appear in bubbles at random screen locations while the front camera captures gaze images, requiring direct foveation for successful memorization. (3) Users recall and input the digit sequence, after which the system validates the input and automatically labels the corresponding gaze data as valid or discards it otherwise.}
  \Description{A three-step workflow illustration for GazeCode. In step 1, a user holds a phone while calibrating device orientation. In step 2, digits appear in small bubbles at different screen positions while the front camera records the user. In step 3, the user enters the recalled digit sequence on a keypad, and the system validates the response to retain or discard the corresponding gaze data.}
  \label{fig:teaser}
\end{teaserfigure}

\maketitle

\section{Introduction}
Front-facing cameras on mobile devices enable appearance-based gaze estimation ``in the wild'', supporting applications~\cite{lei2023end} such as attentive interfaces~\cite{lei2023dynamicread}, privacy~\cite{katsini2020role, lei2023protecting, he2025identity}, and user analytics~\cite{ghosh2023automatic}. However, learning robust gaze estimators requires \emph{large and diverse} datasets of face images paired with accurate ground-truth gaze labels \cite{krafka2016eye}. While lab-based collection provides high fidelity, it is costly and often limited in demographic and environmental diversity~\cite{zhang2020eth}. Crowdsourced, unsupervised collection can scale to thousands of users~\cite{krafka2016eye, valliappan2020accelerating}, but it introduces a central challenge: \textit{how can we verify, without supervision, that a participant actually foveated the on-screen target at capture time?}

This challenge is not merely about missing faces or poor lighting: it is fundamentally about \emph{label validity}. In-the-wild tasks can be (i) \textit{gamed} via guessing (injecting label noise \cite{hsueh2009data}), and (ii) \textit{completed using peripheral vision}, where the participant detects or reads the stimulus without directly looking at the target \cite{stewart2020review,strasburger2011peripheral, lei2023end}. Either behaviour breaks the assumed mapping between the logged target coordinate and the user's gaze, degrading downstream model training.

We introduce \textbf{GazeCode}, a recall-based mobile paradigm for \emph{higher-confidence} in-the-wild gaze labelling. GazeCode combines two complementary mechanisms. (1) \textit{High-entropy verification:} users recall and enter a sequence of $N$ digits, reducing the probability of random success to $P_{\text{guess}}(N)=10^{-N}$ (e.g., $10^{-4}$ for $N{=}4$), making successful guessing negligibly likely. (2) \textit{Anti-peripheral stimulus design:} digits are small, low-contrast, and briefly displayed to reduce parafoveal/peripheral readability, increasing the likelihood that successful recall implies foveation \cite{strasburger2011peripheral,plummer2022effect}. Importantly, the code length $N$ acts as a tunable \emph{verification-entropy knob}: guessing success decreases exponentially with $N$, while per-trial time and memory burden increase approximately linearly, creating a practical entropy--throughput trade off for scalable data collection~\cite{jonides2008mind}. We emphasize that GazeCode focuses strictly on establishing higher-confidence label validity at capture time; evaluating the subsequent impact on downstream gaze estimation accuracy remains a target for future large-scale study.

Our main contributions are: (i) A \emph{recall-based} in-the-wild gaze-label verification paradigm that drastically reduces successful guessing compared to binary-choice validation~\cite{krafka2016eye,hsueh2009data}. (ii) The design and implementation of an Android system that logs synchronized front-camera video, IMU streams, and target events for gaze model training~\cite{yang2022continuous}. (iii) A formative study (N=3) that validates key parameters (opacity, duration) and directly tests peripheral exploitability using a controlled eccentricity manipulation grounded in peripheral-vision literature \cite{strasburger2011peripheral,staugaard2016eccentricity}.

\section{Related Work}
\textit{Collecting gaze datasets.}
Most gaze datasets are captured under supervision with specialized hardware or controlled procedures, yielding high-fidelity labels but constrained diversity~\cite{zhang2020eth}. In contrast, crowdsourced approaches can scale and diversify data sources~\cite{krafka2016eye, valliappan2020accelerating,xu2015turkergaze}, but require robust quality controls to prevent invalid labels.

\textit{In-the-wild gaze collection and its vulnerabilities.}
GazeCapture~\cite{krafka2016eye} pioneered large scale mobile gaze data collection by prompting users to fixate a target and then report a briefly presented letter (L/R) via a binary tap. MPIIGaze~\cite{zhang2017mpiigaze} adopted a related validation strategy on a laptop platform, and similar dot-plus-binary-probe designs have since been widely reused in mobile and RGB-D settings~\cite{lian2019rgbd, arakawa2022rgbdgaze, kellnhofer2019gaze360, lei2025quantifying}. While efficient, these protocols have two practical vulnerabilities for unsupervised deployment: (i) \textit{low-entropy responses} allow a non-trivial probability of correct guessing, admitting mislabeled samples into the dataset; and (ii) \textit{peripheral viewing} can enable users to detect or identify the probe without foveating the logged target, breaking the assumed correspondence between target coordinates and gaze~\cite{strasburger2011peripheral,oderkerk2022fonts}. 

Recent work has therefore focused on strengthening label verification in scalable settings. Yue et al.~\cite{yue2025evaluating} explored more gameful collection tasks to reduce ground-truth errors, while Elfares et al.~\cite{elfares2025qualiteye} proposed privacy-preserving, automated checks for gaze data quality. These efforts reinforced that \emph{verification}--not only scale is a first-order challenge in in-the-wild gaze dataset construction.

\textit{Our position.}
We target a practical gap: a task that (i) is feasible on commodity phones, (ii) offers \emph{stronger on-task evidence} that targets were fixated, and (iii) provides concrete, tunable parameters for balancing usability and label validity. (Note: ``GazeCode'' has also been used as a name for a different eye-tracking tool \cite{benjamins2018gazecode}; our work refers to an in-the-wild mobile data collection paradigm.)

\section{The GazeCode System}
GazeCode is an Android application implemented in Java \cite{gosling2000java} for collecting gaze estimation training data. Each trial produces a short front-camera video and synchronized streams of IMU data and target events.

\vspace{1mm}
\noindent\textit{Design goal:} Successful task completion should be difficult without directly viewing each target at the time it appears.

\begin{figure*}[h]
  \centering
  \includegraphics[width=0.95\linewidth]{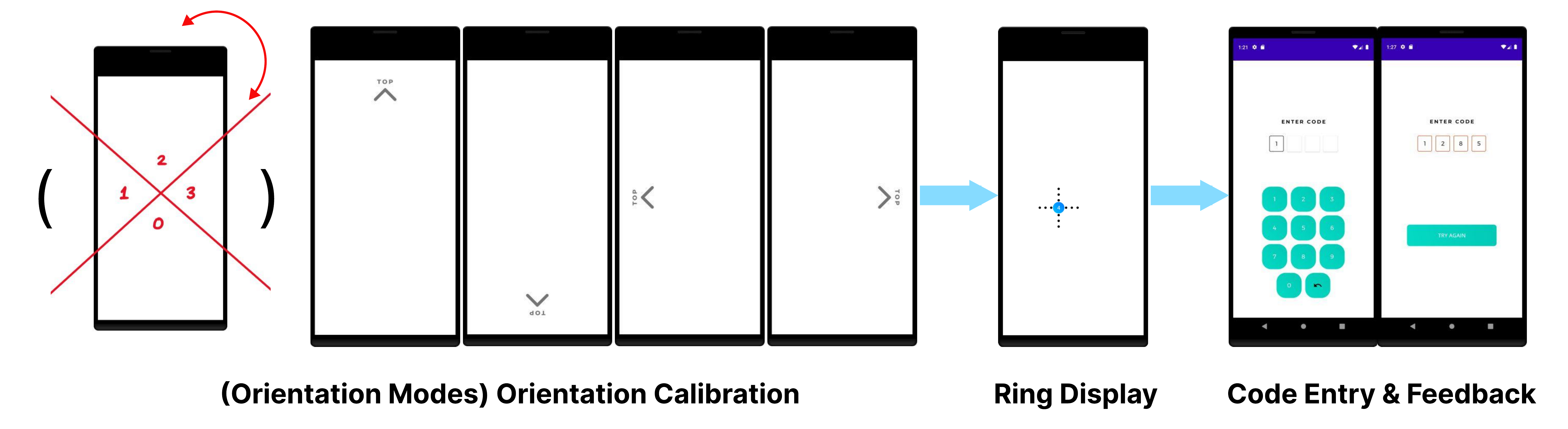}
  \caption{GazeCode flow. (a) Random device orientation is enforced. (b) Digits appear sequentially at random screen locations. (c) The user enters the recalled code. (d) Correct trials are retained as higher-confidence labels.}
  \Description{A screenshot-style flow of the GazeCode app. From left to right: an orientation prompt enforces a required phone pose, the user completes orientation calibration, digits appear sequentially at random screen locations, a ring-based fixation display is shown for the peripheral-reading condition, and the user enters the recalled code and receives correctness feedback. Correct trials are kept as higher-confidence labels.}
  \label{fig:app_flow}
\end{figure*}

\subsection{Core Task: Multi-Digit Recall}
Each trial shows a sequence of $N$ single digits (default $N{=}4$) in ``bubbles'' at pseudo-random screen locations. Each digit is displayed for a fixed duration, then disappears. After the sequence, the user enters the recalled code using a custom keypad.

\begin{itemize}
    \item \textit{Guess resistance:} For digits 0--9, random success probability is $P_{\text{guess}}(N)=10^{-N}$ (e.g., $10^{-4}$ for $N{=}4$), making successful guessing extremely unlikely compared to binary choice validation \cite{hsueh2009data}.
    \item \textit{Operational label check:} Trials with correct code entry are treated as \emph{higher-confidence proxy labels}; incorrect trials are flagged for exclusion or separate analysis. We note that incorrect recall does not necessarily imply incorrect foveation, as failures can also arise from memory or motor entry errors.
\end{itemize}

\paragraph{Entropy--throughput trade off (code length $N$).}
Increasing $N$ improves verification strength by reducing guessing-induced label noise, but it also increases (i) the \emph{time per trial} and (ii) the \emph{memory burden}. Because correct recall acts as an indirect proxy for foveation, conflating perception with memory and motor execution, higher $N$ can introduce \emph{false rejections} (i.e., instances where users foveate correctly but fail recall or mistype the entry). In our prototype workflow, trial time grows approximately linearly with $N$ because digits are presented sequentially and followed by a single entry step, i.e., $T_{\text{trial}}(N)\approx T_{\text{setup}} + N\cdot T_{\text{digit}} + T_{\text{entry}}$, where $T_{\text{digit}}$ subsumes the bubble animation, display interval, and any inter-digit gap. In the formative study, we fix $N{=}4$ as a practical compromise: it reduces random-guess success to $10^{-4}$ while keeping per-trial duration manageable for repeated within-session conditions (Table~\ref{tab:entropy_tradeoff}). This choice prioritizes strong verification entropy without imposing excessive memory burden that could inflate false rejections.

\begin{table}[ht!]
\centering
\small
\begin{tabular}{>{\centering\arraybackslash}p{0.2\linewidth}
                c
                >{\centering\arraybackslash}p{0.4\linewidth}}
\toprule
Code length $N$ & $P_{\text{guess}}(N)=10^{-N}$ & Expected random successes / 1000 trials \\
\midrule
2 & $10^{-2}$ & 10 \\
3 & $10^{-3}$ & 1 \\
4 & $10^{-4}$ & 0.1 \\
5 & $10^{-5}$ & 0.01 \\
\bottomrule
\end{tabular}
\caption{Verification entropy increases with code length: guessing becomes negligible as $N$ grows. However, longer codes increase trial time and memory load, which can reduce throughput and increase false rejections.}
\Description{A three-column table showing the verification entropy trade-off for code length N. Rows list N = 2, 3, 4, and 5, with corresponding random-guess success probabilities of 10^{-2}, 10^{-3}, 10^{-4}, and 10^{-5}, and expected random successes per 1000 trials of 10, 1, 0.1, and 0.01. The table highlights that longer codes greatly reduce guessing but increase trial burden.}
\label{tab:entropy_tradeoff}
\end{table}

\subsection{Anti-Peripheral Stimulus Design}
Peripheral vision has reduced acuity and contrast sensitivity, especially for fine symbols \cite{strasburger2011peripheral,venkataraman2017peripheral}. GazeCode exploits this by making digits harder to identify outside foveal vision:
\begin{itemize}
    \item \textit{Small targets:} Bubbles/digits are kept small while remaining legible.
    \item \textit{Low contrast:} Digit opacity is reduced (e.g., 0.1), increasing the likelihood that correct reading requires foveation \cite{plummer2022effect}.
    \item \textit{Brief presentation:} Digits can be shown briefly (e.g., 150ms), limiting time for gaze shifts.
\end{itemize}
Because correct recall could still fail due to memory/input errors, we additionally validate peripheral exploitability directly in our study (\S\ref{sec:formative}).

\subsection{Encouraging Pose and Gaze Diversity via Orientation Gating}
To diversify relative gaze angles around the front camera, each trial begins with a randomly selected device orientation mode: portrait, reverse portrait, landscape, or reverse landscape. The task starts only once sensors confirm the required mode is held. This produces target distributions that surround the camera in multiple directions (rather than concentrating in one region), complementing existing mobile protocols \cite{krafka2016eye,lei2025mac}.

\subsection{Synchronized Data Logging}
In each trial, GazeCode records: \textit{Front-facing camera video:} 640 $\times$ 480 resolution, 16 frames per second (fps), MP4 format \cite{yang2022continuous}. \textit{Inertial measurement unit (IMU) data stream:} Accelerometer and gyroscope (x,y,z axes) readings for device motion. \textit{Target events:} The (x, y) target coordinates and appearance/disappearance timestamps for each digit. All streams use high-resolution timestamps (nanoseconds) to support accurate alignment between video frames and target events \cite{burger2018synchronizing}. To reduce runtime overhead, the app does not perform on-device face detection; face/eye detection and frame filtering are intended as post-processing steps.

All streams use high resolution timestamps (nanoseconds) to support alignment between video frames and target events \cite{burger2018synchronizing}. To reduce runtime overhead, the app does not perform on device face detection; face/eye detection and frame filtering are intended as post processing.

\begin{table}[h]
\centering
\small
\begin{tabular}{ll}
\toprule
Code length $N$ & 4 digits \\
Digit opacity $\alpha$ & 1.0 (visible), 0.1 (faint) \\
Default duration & 300ms (CONTROL); 50/150/300ms (INTERVAL) \\
RING radius & 0.13 / 0.23 / 0.33 inches \\
Video logging & 640$\times$480 @ 16 fps (front camera) \\
Sensors & accelerometer + gyroscope \\
\bottomrule
\end{tabular}
\caption{Key parameters used in the current prototype and study.}
\Description{A parameter summary table for the GazeCode prototype and formative study. It lists a 4-digit code length, digit opacity levels of 1.0 and 0.1, presentation durations of 300 ms for CONTROL and 50/150/300 ms for INTERVAL, ring radii of 0.13, 0.23, and 0.33 inches, front-camera video logging at 640 by 480 pixels and 16 frames per second, and accelerometer plus gyroscope sensor streams.}
\label{tab:params}
\end{table}

\section{Formative Evaluation}
\label{sec:formative}
We conducted an in-person formative study with three participants (2 male, 1 female, ages 20--22), including the primary author, using a single Android device. The study was designed to provide \emph{indicative feasibility evidence} and parameter sensitivity analysis, rather than generalizable performance estimates, for the proposed verification paradigm. Specifically, our goals were to (i) determine usable ranges for digit opacity and presentation duration, and (ii) test whether low-opacity digits meaningfully reduce peripheral exploitability.

\subsection{Design}
The study used a within subjects design. Participants completed six repeat trials per condition; each trial required recalling a 4-digit code. Independent variables and experiments:

\begin{enumerate}
    \item \textit{CONTROL (300ms):} baseline recall accuracy with fixed duration.
    \item \textit{TAP:} digit remains until user taps; provides an \emph{upper bound} on time-to-read+respond (includes motor latency) \cite{han1970reaction,derosa1970recognition}.
    \item \textit{RING:} four fixation dots appear on a ring around the bubble. Participants fixate a dot while attempting to read the central digit. Ring radius is small/medium/large, directly manipulating eccentricity \cite{strasburger2011peripheral,staugaard2016eccentricity}.
    \item \textit{INTERVAL:} digit duration varies (50ms, 150ms, 300ms) to find feasible lower bounds for attentive reading.
\end{enumerate}

\subsection{Measures and Analysis}
We report (i) \textit{code success rate} (correctly entered codes / trials) as a task-level indicator of successful perception+memory+entry, and (ii) \textit{tap latency} in TAP as an operational upper bound on read time. For TAP timing, we only analyze trials where the final code is correct (to avoid counting non-attentive trials). In this formative prototype, the primary reported outcome is trial-level correctness; we do not yet decompose failures into perceptual, memory-maintenance, and motor-entry components, which we identify as an important next step for quantifying false-rejection trade-offs.

\section{Results}
\paragraph{CONTROL: Baseline Readability}
With fully visible digits ($\alpha{=}1.0$), all participants achieved 100\% success. With faint digits ($\alpha{=}0.1$), mean success decreased to 67\%, indicating increased perceptual difficulty and/or concentration demands.

\begin{figure*}[t]
  \centering
  \captionsetup{font=small}
  \begin{subfigure}{0.32\textwidth}
    \includegraphics[width=\linewidth]{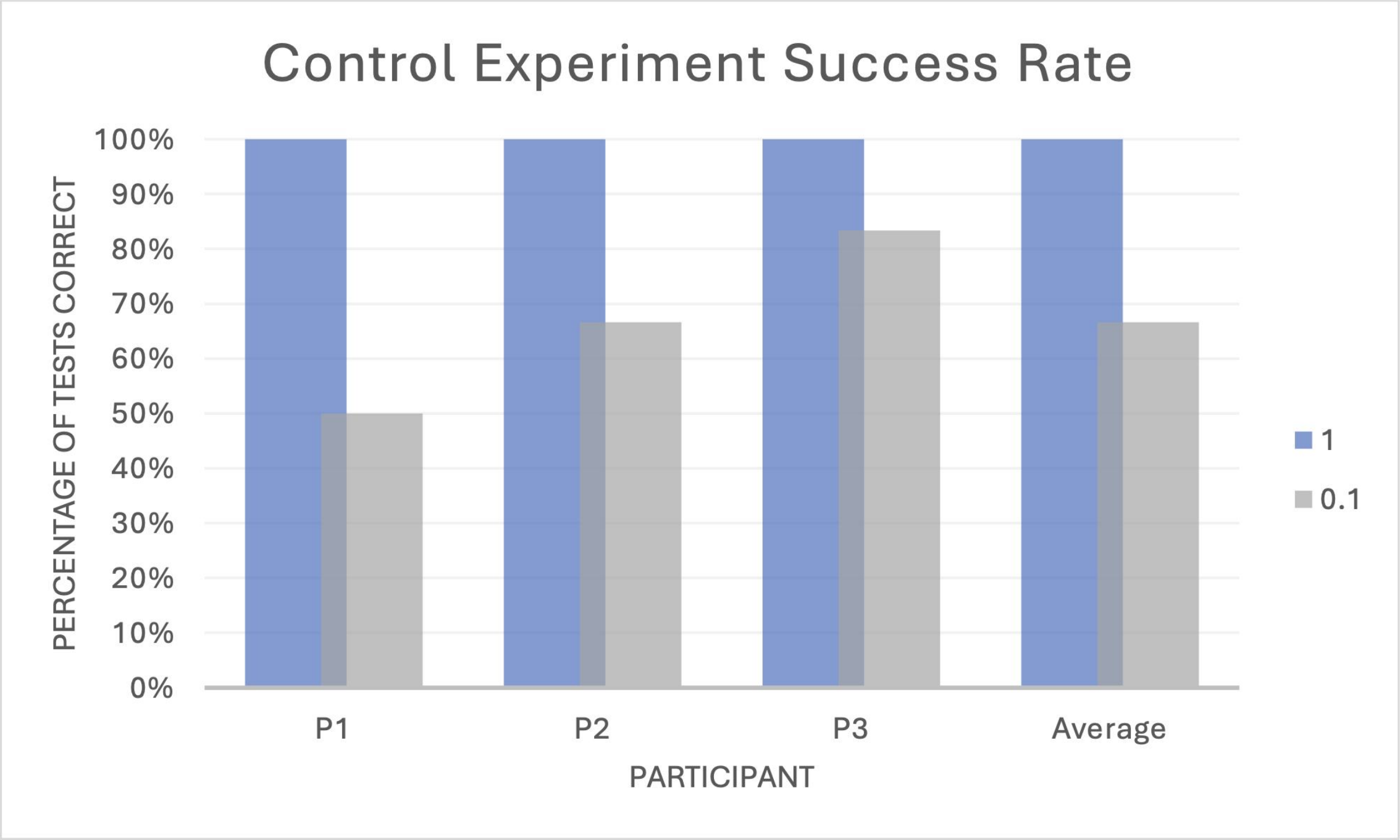}
    \caption{CONTROL: Faint digits ($\alpha=0.1$) lower success rates, indicating higher difficulty.}
    \label{fig:control}
    \Description{A bar chart of CONTROL task success rates comparing visible digits (opacity 1.0) and faint digits (opacity 0.1) across participants and the average. Visible digits achieve uniformly high success, while faint digits show lower and more variable success, indicating increased task difficulty.}
  \end{subfigure}\hfill
  \begin{subfigure}{0.32\textwidth}
    \includegraphics[width=\linewidth]{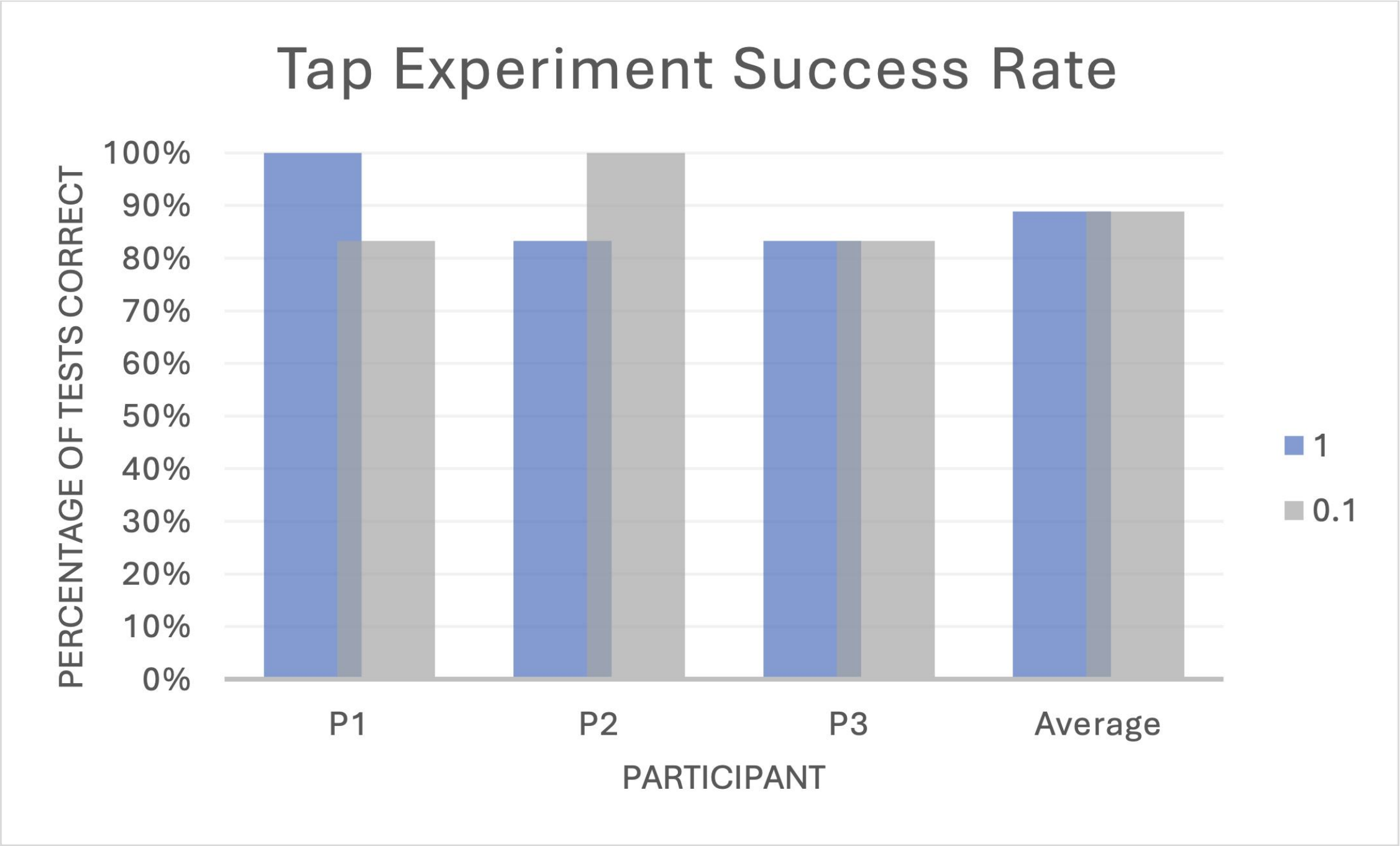}
    \caption{TAP: Reaction times increase by $\approx$60\% for faint digits.}
    \label{fig:tap}
    \Description{A bar chart for the TAP condition comparing reading-and-response time for visible versus faint digits. Faint digits require longer time-to-tap than visible digits across participants, and the average time is substantially higher for faint digits, indicating increased perceptual effort.}
  \end{subfigure}\hfill
  \begin{subfigure}{0.32\textwidth}
    \includegraphics[width=\linewidth]{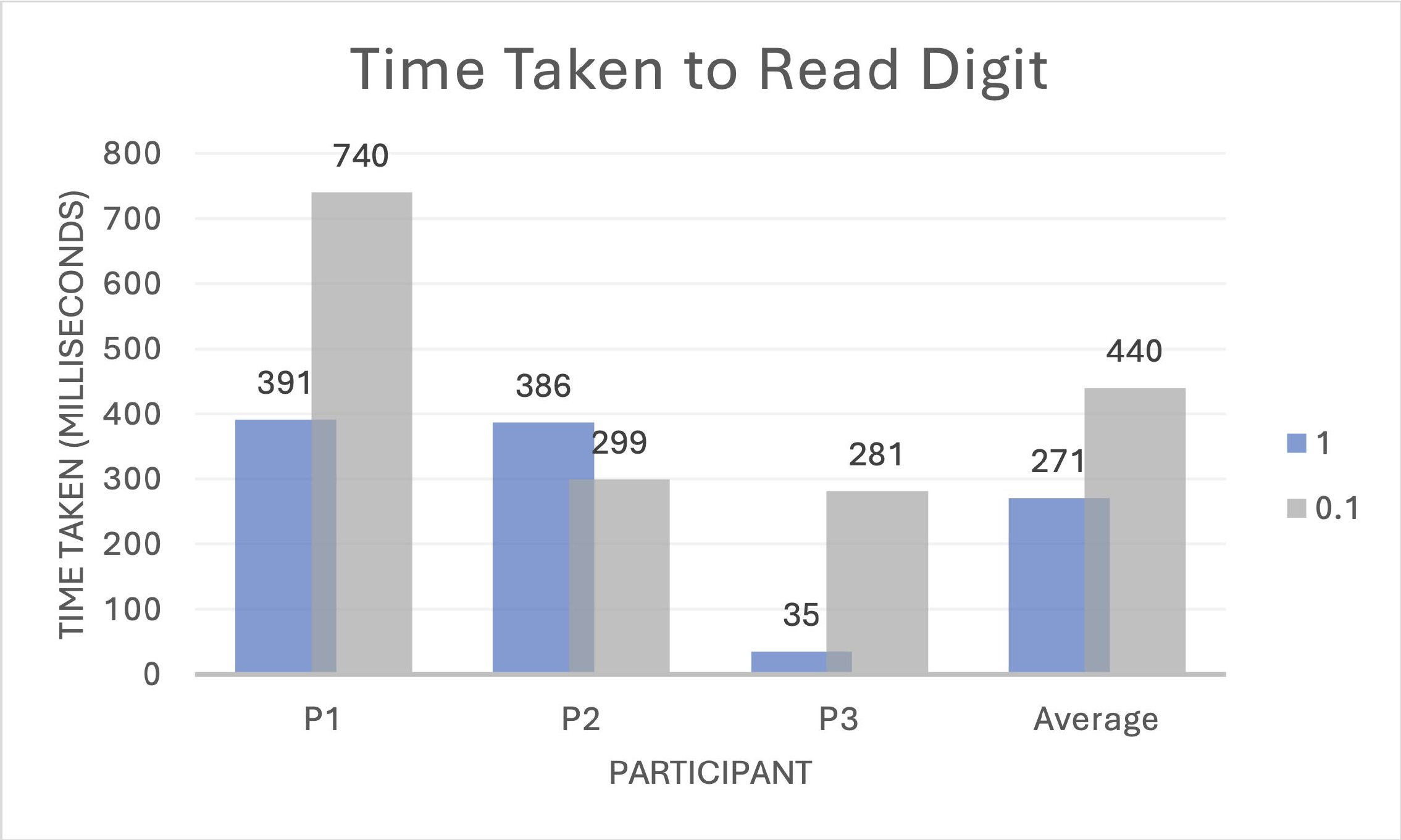} 
    \caption{RING: Faint digits become unreadable at 0.33 inches peripheral distance.}
    \label{fig:time}
    \Description{A bar chart for the RING condition showing peripheral readability performance as ring radius increases. For visible digits, performance remains relatively high at larger eccentricities. For faint digits, success drops sharply with increasing ring radius, showing that low-opacity digits are difficult to identify without direct foveation.}
  \end{subfigure}
  \vspace{-2mm}
  \caption{Formative results. Low opacity (0.1) significantly increases the time required to read digits and drastically reduces peripheral legibility, enforcing foveation.}
  \Description{A three-panel results figure summarizing the formative evaluation. Panel (a) shows lower success for faint digits in the CONTROL condition. Panel (b) shows longer reading-and-response times for faint digits in the TAP condition. Panel (c) shows reduced peripheral readability for faint digits in the RING condition as eccentricity increases. Together, the panels indicate that low-opacity digits increase difficulty and reduce peripheral exploitability, reinforcing foveation.}
  \label{fig:results_overview}
\end{figure*}

\paragraph{TAP: Upper Bound on Read Time}
Across participants, mean time-to-tap was 271ms for visible digits and 440ms for faint digits, suggesting that low-opacity digits increase processing time and that 300ms is a conservative default. As expected, TAP includes motor and device latency and therefore upper-bounds perceptual time \cite{taylor2006noisy}.

\paragraph{RING: Peripheral Exploitability}
RING provides the most direct evidence for foveation enforcement. For visible digits ($\alpha{=}1.0$), participants often remained accurate even at larger ring radii, consistent with the ability to identify high contrast symbols in parafovea/periphery \cite{oderkerk2022fonts}. In contrast, for faint digits ($\alpha{=}0.1$), performance collapsed as eccentricity increased: mean success dropped from 33\% (small ring) to 11\% (large ring), and two participants could not reliably read faint digits while fixating any ring distance. This pattern supports our core design rationale: \textit{low-opacity digits substantially reduce the viability of peripheral reading} under the tested conditions, thereby strengthening the inference that correct recall is associated with target foveation.

\paragraph{INTERVAL: Minimum Feasible Duration}
Visible digits remained highly readable even at 50ms, suggesting 300ms is generous for high-contrast stimuli. For faint digits, participants still achieved $>{}60\%$ mean success at 50ms, but performance was more variable. Taken together with TAP, these results suggest that \textit{short durations (e.g., 150ms) with low opacity} can balance usability with foveation pressure.

\section{Discussion and Implications}
This work aims to strengthen \emph{label validity verification} in unsupervised gaze data collection by combining (i) high-entropy verification (multi-digit recall) with (ii) \emph{anti-peripheral} stimulus design (faint, small, and brief digits). The RING manipulation provides a targeted validation of the second mechanism, showing that low-opacity digits are substantially less readable even at modest eccentricities.

Based on these findings, we summarize four key recommendations for implementing in-the-wild data collection:

\noindent 1. \textbf{Verification entropy:} Prefer multi-step recall (e.g., 4 digits) over binary probes to reduce guessing-induced label noise.

\noindent 2. \textbf{Anti-peripheral stimuli:} Use low contrast (e.g., $\alpha \approx 0.1$) to reduce peripheral exploitability \cite{strasburger2011peripheral,plummer2022effect}.

\noindent 3. \textbf{Brief presentation:} Use short durations (e.g., 150\,ms) to limit time for gaze shifts while retaining usability.

\noindent 4. \textbf{Pose diversity via orientation gating:} Enforce multiple device orientations to better surround the camera with labelled targets.


\subsection{Limitations and Future Work}
This is a formative, small-scale study (N=3) on a single device, so the results should be interpreted as indicative rather than generalizable. In particular, we do not yet characterize robustness under real-world motion, lighting variation, screen-size differences, or broader hardware diversity. Next, we will deploy at scale (e.g., crowdsourcing) to test robustness across devices, environments, gestures~\cite{lei2026people}, and demographics~\cite{xu2015turkergaze}. Our validity signal is currently \emph{trial-level} correctness, which conflates perception, memory, and motor entry; longer codes can raise false rejections even when foveation is correct. Future versions should decouple these factors via partial-report, per-digit confidence, or lightweight intermediate checks \cite{damiano2019distinct}. We also plan to quantify the usability cost of orientation gating (e.g., compliance time, abandonment, and perceived burden) while preserving its pose-diversification benefit. Finally, we will validate downstream utility by training a gaze estimator (e.g., ViT-based models~\cite{cheng2024lightweight,lei2025mac}) and benchmarking against established in-the-wild protocols, with ablations over $N$, opacity, and duration~\cite{krafka2016eye}.

\section{Conclusion}
We introduced GazeCode, a recall-based in-the-wild gaze data collection paradigm that strengthens label validity through multi-digit recall verification and anti-peripheral stimulus design. A formative study shows that low-opacity digits dramatically reduce peripheral readability under controlled eccentricity manipulations, supporting the inference that correct trials more likely reflect target foveation. These findings provide concrete, tunable guidelines for strengthening label-validity verification in large-scale mobile gaze data collection and for designing future deployments that balance verification strength, usability, and throughput.


\begin{acks}
We thank all participants for their time and contributions. We also thank the CHI reviewers for their constructive feedback, which helped improve the clarity and presentation of this work. Finally, we are grateful to our colleagues for helpful discussions and support throughout the project.
\end{acks}

\bibliographystyle{ACM-Reference-Format}
\bibliography{reference}


\clearpage
\appendix

\end{document}